\documentclass[%
twocolumn,
 amsmath,amssymb,
 prl,
]{revtex4-1}

\usepackage{xcolor}
\usepackage{amsmath}
\usepackage{graphicx}
\usepackage{dcolumn}
\usepackage{bm}

\usepackage[polish,english]{babel}
\usepackage[T1]{fontenc}
\usepackage[utf8]{inputenc}
\usepackage{lmodern}
\usepackage{braket}

\usepackage[pdftex,bookmarks=true,bookmarksopen,bookmarksnumbered,
                colorlinks,
                linkcolor=black,
                citecolor=black,
                urlcolor=black]{hyperref}

\renewcommand{\figurename}{Figure}
\makeatletter
\renewcommand*{\fnum@figure}{{\normalfont\bfseries \figurename~\thefigure}}
\renewcommand*{\@caption@fignum@sep}{ $|~$}

\begin{document}

\title{12-spin-qubit arrays fabricated on a 300 mm semiconductor manufacturing line}
\newcommand\Intel{Intel Corporation, Technology Research Group, Hillsboro, OR 97124, USA}


\author{Hubert C. George$^{1,2}$}
\author{Mateusz T. M\k{a}dzik$^1$}
\author{Eric M. Henry$^1$}
\author{Andrew J. Wagner$^1$}
\author{Mohammad M. Islam$^1$}
\author{Felix Borjans$^1$}
\author{Elliot J. Connors$^1$}
\author{J. Corrigan$^1$}
\author{Matthew Curry$^1$}
\author{Michael K. Harper$^1$}
\author{Daniel Keith$^1$}
\author{Lester Lampert$^1$}
\author{Florian Luthi$^1$}
\author{Fahd A. Mohiyaddin$^1$}
\author{Sandra Murcia$^1$}
\author{Rohit Nair$^1$}
\author{Rambert Nahm$^1$}
\author{Aditi Nethwewala$^1$}
\author{Samuel Neyens$^1$}
\author{Bishnu Patra$^1$}
\author{Roy D. Raharjo$^1$}
\author{Carly Rogan$^1$}
\author{Rostyslav Savytskyy$^1$}
\author{Thomas F. Watson$^1$}
\author{Josh Ziegler$^1$}
\author{Otto K. Zietz$^1$}
\author{Stefano Pellerano$^1$}
\author{Ravi Pillarisetty$^1$}
\author{Nathaniel C. Bishop$^1$}
\author{Stephanie A. Bojarski$^1$}
\author{Jeanette Roberts$^1$}
\author{James S. Clarke$^1$}
\affiliation{$^1$\Intel}
\affiliation{$^2$hubert.c.george@intel.com}

\begin{abstract}
   \textbf{ABSTRACT: }
   \textbf{Intel’s efforts to build a practical quantum computer are focused on developing a scalable spin-qubit platform leveraging industrial high-volume semiconductor manufacturing expertise and 300 mm fabrication infrastructure. Here, we provide an overview of the design, fabrication, and demonstration of a new customized quantum test chip, which contains 12-quantum-dot spin-qubit linear arrays, code named Tunnel Falls. These devices are fabricated using immersion and extreme ultraviolet lithography (EUV), along with other standard high-volume manufacturing (HVM) processes, as well as production-level process control. We present key device features and fabrication details, as well as qubit characterization results confirming device functionality. These results corroborate our fabrication methods and are a crucial step towards scaling of extensible 2D qubit array schemes.\newline \newline
   }
\end{abstract}

\date{\today}
             
\maketitle

Electron spin-qubit devices based on gate-defined silicon quantum dots (QD) are a promising platform for quantum computing due to their small device size, scalability, long spin coherence times \cite{veldhorst2014addressable,vandersypen2019quantum, zajac2015reconfigurable, borselli2011pauli} and high single-qubit and two-qubit fidelities \cite{yoneda2018quantum, mills2022two, xue2022quantum}. Their inherent compatibility with complementary metal-oxide-semiconductor (CMOS) processing \cite{auth201710nm,zwerver2022qubits,koch2024industrial300mmwaferprocessed} is especially attractive; however most spin qubit devices are fabricated using a combination of electron-beam lithography (EBL) and lift-off  processes to create overlapping gate structures \cite{angus2007gate, zajac2016scalable}, which are not utilized in modern CMOS fabrication flows. A more scalable approach that still incorporates EBL, but replaces lift-off with etching processes \cite{ha2021flexible} has successfully been used to demonstrate  high-fidelity qubit operations \cite{weinstein2023universal}. Other approaches employ modified fin- and planar-based CMOS devices \cite{vinet2018towards, elsayed2023comprehensive}. 

Recently, highly uniform and reproducible linear arrays of quantum dots have been demonstrated in a planar Si/SiGe technology using a 300 mm manufacturing infrastructure  \cite{neyens2024probing}. Here, we present an overview of the fabrication process for the quantum chip used in the experiment mentioned above. Additionally, we provide evidence for the success of this process by showcasing results from a tune up of a 12-quantum-dot-qubit device, which represents the largest reported qubit count in a single spin-qubit device to date.

\begin{figure}[tb]
	\includegraphics[width=\columnwidth]{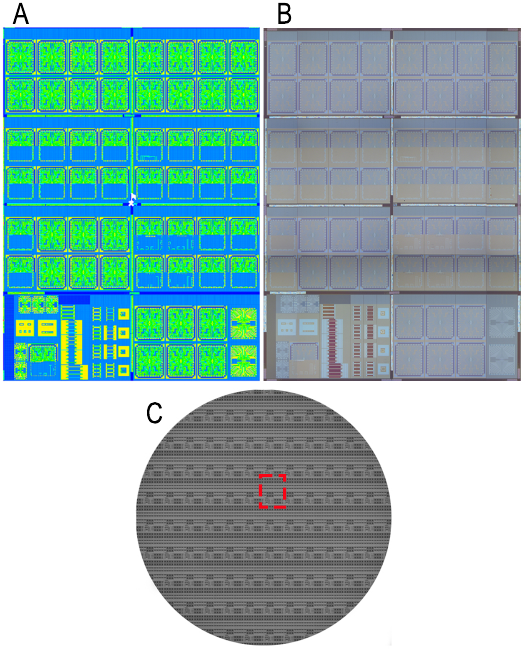}
	\caption{\textbf{a)} Test chip (TC) floor plan layout. 8 seats with 22 12QD and 21 3QD devices with different design layout variations. In addition, transistors, HB and other test devices are part of the TC infrastructure \textbf{b)} Optical image of the TC after fabrication. \textbf{c)} In-line optical image of full wafer after fabrication. Highlighted in red is one full TC die.}
	\label{fig:TC}
\end{figure}

The Tunnel Falls test chip (TC) contains two primary qubit device types, the 12QD and 3QD devices, with design layout variations and gate pitches ranging from 45 to 100 nm (gate to gate). Additionally, we include a wide range of  structures to aid process development. Transistor and Hall bar (HB) devices are used for general material characterization, both at room and low temperatures. From them, we extract metrics such as mobility, threshold voltage, subthreshold slope, and interface-trap density, which are used to track material quality, and assist with overall process assessment and material optimization. Lithography and process-calibration devices are also included within the TC device infrastructure for further process characterization and troubleshooting. Fig.\ref{fig:TC}a shows the Tunnel Falls quantum TC floorplan with a corresponding optical image of the full reticle (repeated pattern on a wafer) taken after fabrication (Fig.\ref{fig:TC}b). Altogether, there are more than 24,000 testable quantum dot devices on every wafer (Fig.\ref{fig:TC}c), including 1,230 variations of the 12QD device. 

During the design phase, we employ established CMOS production design rules required for a reliable fabrication process. Such rules dictate device feature size and dimensions for every layer, as well as interactions between layers. TC layouts undergo the same validation flows and checks used by transistor products during the tape-out process, and include the corresponding optical proximity correction (OPC) adjustments needed for pattern fidelity \cite{venkatesan2023direct}. In addition, the TC layouts incorporate the required extra fill features/structures which are used to maintain pattern uniformity within the local device area and across the entire wafer.  The resulting device functionality, uniformity and reproducibility \cite{neyens2024probing} reflect the robustness and integrity of the design and the fabrication process.

\begin{figure*}[t]
	\includegraphics[width=\textwidth]{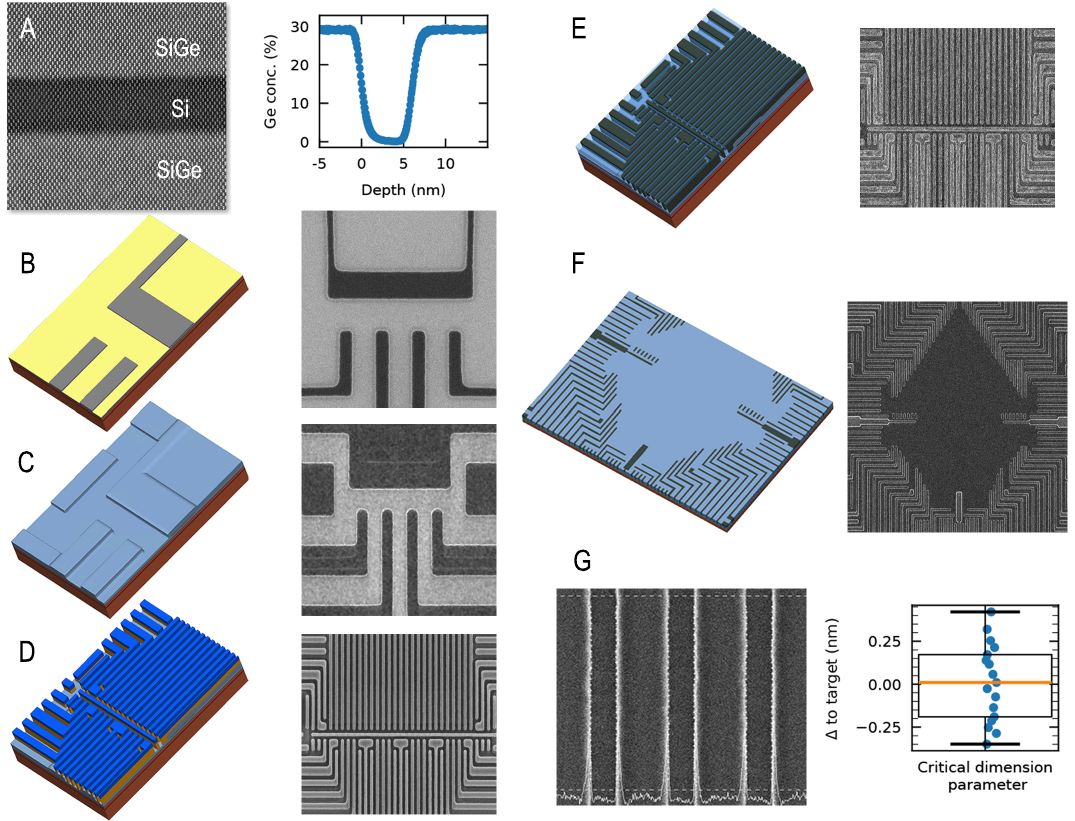}
	\caption{\textbf{a)} STEM cross-section and SIMS data of the Si/SiGe heterostructure stack. 12QD fabrication flow schematic details after \textbf{b)} screening gate (SG) patterning,\textbf{ c)} qubit/SET trench etch, \textbf{d)} gate patterning, \textbf{e)} post-RMG process, and \textbf{f)} after contact metallization/CMP. \textbf{ g)} Top-down SEM metrology structure image with corresponding CD measurements.} 
	\label{fig:fab}
\end{figure*}

An overview of the fabrication process flow of Tunnel Falls devices can be seen in Fig.\ref{fig:fab}. We first epitaxially grow a thick strain relaxed buffer (SRB) of linearly graded Si$_{0.7}$Ge$_{0.3}$ on a base Si substrate. The Si/SiGe heterostructure is grown on top of the SRB, with a strained Si quantum well (QW) of thickness between 3 nm and 10 nm. The Si QW is then covered by a 30-75 nm Si$_{0.7}$Ge$_{0.3}$ barrier layer and capped with a thin layer of Si (1-2 nm). The SiGe barrier separates the QW and therefore the electrons from the disordered semiconductor-oxide interface. High resolution transmission-electron-microscope (TEM) image of the Si/SiGe heterostructure can be seen in Fig.\ref{fig:fab}a. Once the Si/SiGe heterostructure has been grown, the gate-oxide stack, consisting of SiO$_2$ (5-10 nm) and a HfO$_2$ (high-k) material (5-10 nm) are deposited. To form the screening gates (SG), a thin layer of metal (TiN) is deposited, patterned using immersion lithography, and etched selective to the gate dielectric stack (Fig.\ref{fig:fab}b). Then, a layer of SiO$_2$ dielectric material is deposited, patterned using immersion lithography, and etched to form trenches between the buried screening gates (SG) (Fig.\ref{fig:fab}c). Ohmic regions are lithographically defined, and phosphorus dopants are implanted in the localized contact areas and subsequently annealed. The qubit, SET, and center-screening (CS) gates are defined by a single pass of extreme ultraviolet lithography (EUV) patterning using industry-standard replacement metal gate (RMG) process flow \cite{natarajan200832nm,mistry200745nm,Auth200845nm}. The RMG process involves using a sacrificial material that is deposited and polished using chemical mechanical polishing (CMP). The material stack is then EUV patterned, as shown in Fig.\ref{fig:fab}d. SiO$_2$ is deposited and CMP polished, exposing the sacrificial material, which is etched away and then replaced by the gate oxides, work-function and metal fills. A last metal CMP is used to define the final gate structures (Fig.\ref{fig:fab}e). The RMG fabrication process enables the patterning of highly-uniform gates with tight pitches, using fully compatible CMOS transistor gate manufacturing.

The combination of the SiO$_2$ oxide trenches and the EUV patterning creates “via-like” plunger/barrier gate structures. The cointegration of the SG and plunger/barrier gates, shown in the 3D device schematic of Fig.\ref{fig:fab}e, allows full electrostatic control of an otherwise planar heterostructure, in contrast to Intel's previously reported fin- and gate-defined qubit devices \cite{zwerver2022qubits,pillarisetty2018qubit}. To preserve the quality of the interfacial gate oxide stack throughout the process flow, the RMG process was optimized for etch selectivity, thermal budget, material stack and gate profile. For instance, power bias adjustments were made on dry etch processes to minimize plasma and sputtering damage, thereby protecting the underlying oxide films that form part of the device's gate oxide.

After the metal-gate processing is completed, a single contact interconnect layer is patterned. This involves depositing SiO$_2$ as the interlayer dielectric material (ILD) then etching trench and via structures, which will be used to route connections from the bonding pads to the gate lines and ohmic contacts. These contact line structures are filled with low resistance metal (Cu), and then CMP polished to completely isolate the final interconnect lines (Fig.\ref{fig:fab}f). 

Throughout the entire fabrication process we utilize production-level process control \cite{sarfaty2002advance} and in-line metrology \cite{orji2018metrology}. Wafers are closely monitored during fabrication, and in-line data is constantly collected after processing. At every patterning step, key structures are inspected using high-resolution optical and e-beam imaging, and validated to ensure their dimensions are within well-established statistical process control (SPC) limits, which are determined by methodologies such as Six Sigma \cite{montgomery2008overview}. Fig.\ref{fig:fab}g shows an in-line top-down SEM image of one of the metrology structures after litho patterning, along with the distribution of wafer measurements collected over a period of $\sim$8 weeks. It is worth noting that each data point represents the mean of multiple within-wafer measurements, all of which deviate by less than 0.5 nm from the target CD (critical dimension). The implementation of process control techniques has led to consistent spin-qubit-device fabrication, producing highly uniform arrays, with 96\% of QD devices successfully tuned to the single-electron regime using automated routines across full 300 mm wafers, measured in a cryo-prober operating at temperatures of 1.7 K \cite{neyens2024probing}. This manufacturing method and approach ensures high process quality, statistical uniformity and reproducibility, as well as the required process variation reduction that is necessary for device scalability.

\begin{figure}[tb]
	\includegraphics[width=\columnwidth]{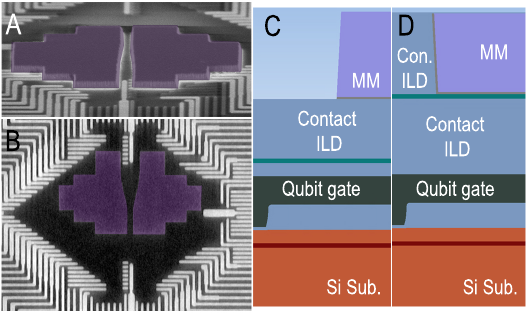}
	\caption{\textbf{a)} SEM (tilted-view) of a coupon-lab line micromagnet (MM) after lift-off. \textbf{b)} SEM (top-view) of an in-line MM after fabrication. Device schematics showing MM locations respect to QW for \textbf{c)} coupon-lab line and \textbf{d)} in-line fabrications.} 
	\label{fig:MM}
\end{figure}

Devices that are intended for resonantly driven spin-qubit measurements \cite{loss1998quantum} are further processed by patterning micro-magnets (MMs) on top, either by using the coupon-lab (using diced samples) or the full wafer lines. For the coupon-lab fabrication, cobalt MMs are patterned on top of the contact ILD layer using e-beam lithography and lift-off processes \cite{yoneda2015robust}, as shown in Fig.\ref{fig:MM}a/c. For the fully integrated process, MMs are fabricated within the contact ILD material, just on top of the plunger/barrier gates. Its patterning process starts by etching the MM trenches on the SiO$_2$ contact ILD, then filling the structures with a metal liner (Ti) and a metal seed (Co), followed by the deposition of bulk cobalt fill material. MMs are fully defined after a CMP polish process step. Fig.\ref{fig:MM}b shows a top SEM image of the integrated MM process after fabrication. Some of the benefits of the integrated MM process is that it allows to pattern the magnet structures closer to the QW (Fig.\ref{fig:MM}d), providing stronger driving gradients for qubit control and addressability, with minimal dephasing \cite{dumoulin2021low}. In addition, samples can be taken directly from wafer-level cryo-prober measurements to single-die-level dilution refrigerator measurements without the need for additional fabrication steps.  

\begin{figure}[tb]
	\includegraphics[width=\columnwidth]{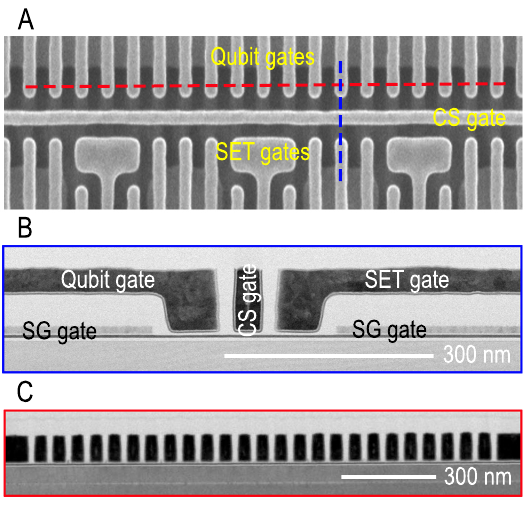}
	\caption{\textbf{a)} In-line scanning electron micrograph SEM (top view) of the 12QD qubit/SET gates after EUV patterning. TEM cross-sections of 12QD device after fabrication \textbf{b)} along the qubit/SET gates, and \textbf{c)} across the qubit array gates.} 
	\label{fig:TEMs}
\end{figure}

The primary 12QD qubit device (Fig.\ref{fig:TEMs}a) consists of a linear array of 25 consecutive gates. Plunger gates are above QDs to control their dot occupation/electro-chemical potential and barrier gates are used for lateral confinement and tunnel coupling control of adjacent QDs. Additional 4 semi-independent single-electron-transistor (SET) detectors are used for charge sensing and qubit readout. The qubit array and the SET sensors are separated by a center-screening gate (CS) that is used for improved electrostatic control and for spin manipulation via electric dipole spin resonance (EDSR) \cite{nowack2007coherent,pioro2008electrically}, as shown in Fig.\ref{fig:TEMs}a. The TEM images from Fig.\ref{fig:TEMs}b-c show cross-sections of a 12QD device along the qubit/SET gate direction and across the qubit array gates, respectively. Side-screening gate structures (SG) have been integrated to enhance the electrostatic control and confinement of the QDs. These SG gates are patterned underneath the qubit/SET gates, embedded within the dielectric material (Fig.\ref{fig:TEMs}b) behind the qubit and SET gates, respectively.

\begin{figure}[!tb]
	\includegraphics[width=\columnwidth]{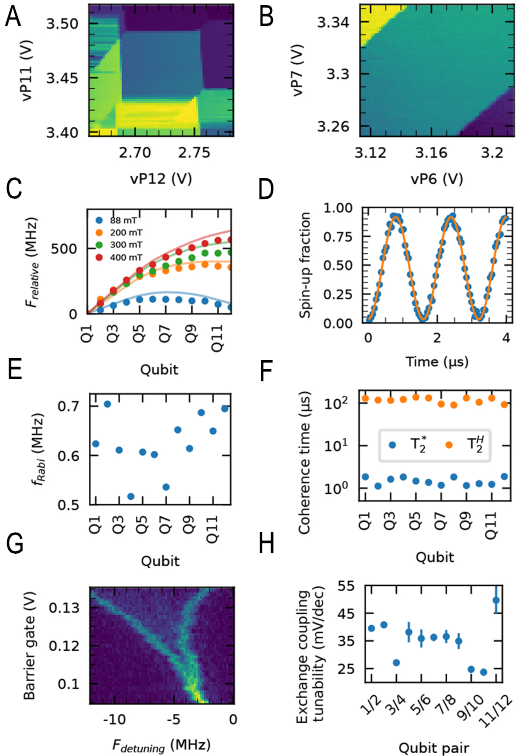}
	\caption{Charge stability diagrams of \textbf{a)} external quantum dot pair with access to reservoir and \textbf{b)} internal quantum dot pair.  \textbf{c)} Qubits resonant frequency separation, as measured by relative shift of frequency of qubits within an array, compared to qubit housed in a left most quantum dot (Q1). The magnetic field gradient of the on chip MM changes in a function of the externally applied magnetic field $B_\text{ext}$. We observe that at $B_\text{ext} = $ 400 mT, the minimum frequency spacing between qubits is 14.8 MHz. \textbf{d)} Rabi oscillation of qubit 1 driven by MW signal applied to central screening gate at qubits resonant frequency. \textbf{e)} A comparison of frequency of Rabi oscillation driven at the same MW output power, and \textbf{f)} pure dephasing time T$_2^*$ and T$_2^\text{H}$ Hahn echo times, for of the entire 12QD device. \textbf{g)} Exchange coupling spectroscopy measurement, where the splitting of the qubit resonance frequency in function of barrier voltage is proportional to the exchange coupling between qubits. \textbf{h)} Exchange coupling tunability extracted for each quantum dot pairs. The average pulse amplitude needed for an order of magnitude change in exchange coupling is $\sim$ 35 +/- 7.2 mV/dec.}
	\label{fig:12Q}
\end{figure}

The final verification of the Tunnel Falls device is performed in a dilution refrigerator at base temperature of 20 mK. We tune the 12QD device (with coupon-lab MM) to the (3,1,3,1,3,1,3,1,3,1,3,1) charge state and keep it in the same DC tuning for all the following measurements. To achieve this electron occupation, we use an arbitrary waveform generator to generate sweep and step signals on calibrated virtual plunger gates \cite{mills2019shuttling,hsiao2020efficient} of two neighboring quantum dots, while observing the current changes through the nearby SET, creating charge stability diagrams. We run these measurements in a loop to create a live image and observe abrupt changes of the SET current which correspond to loading/unloading electrons on quantum dots. We first form a double dot at the edge of the quantum dot array by appropriately adjusting voltages on quantum dot gates (Fig.\ref{fig:12Q}a). Then, we propagate the average voltage of the two plunger gates and of the barrier gate in between the two quantum dots, to the rest of the plunger and barrier gates in the array, respectively. This serves as the starting point, made possible by the high uniformity of threshold voltages across Tunnel Falls devices \cite{neyens2024probing}, resulting in 12 QDs being formed with few electrons in each. Subsequently, we park the plunger gate voltage of the outer double quantum dot on their avoided crossing to facilitate fast electron tunneling to inner quantum dots, via co-tunneling process. This allows us to view the inner charge stability diagrams similarly to outer quantum dots, with all loading lines (Fig.\ref{fig:12Q}a), as opposed to just seeing inter-dot tunneling lines in a double dot with fixed number of electron due to isolation from the electron reservoir (Fig.\ref{fig:12Q}b). This is necessary for deterministic electron counting. We follow this procedure from both edges of the quantum dot array, to finally set the count of middle quantum dots (QD6-QD7). Then moving outward, we count electrons and set the plunger gate voltages in the center of the desired charge cells, reaching the desired charge occupation.

The (3,1,3,1,3,1,3,1,3,1,3,1) charge configuration was chosen to enable Pauli spin blockade (PSB) readout in between neighboring qubits (defined on a single spin of electrons trapped in quantum dots \cite{loss1998quantum}), without the restriction of valley splitting on the PSB-readout window \cite{philips2022universal}. While valley splittings weren't directly measured in this device, we probe them in similar devices from the same wafer. We find a mean valley splitting of 78$\mu$eV. The readout window size, and consequentially the pulse calibration error margin, is proportional to the singlet-triplet (ST) splitting in a quantum dot. In (2,0), the ST splitting is often limited by the valley splitting, whereas in (4,0) the ST splitting is largely determined by the larger orbital splitting due to electron shell filling \cite{leon2020coherent}, resulting in (4,0) having a larger window. The readout scheme operates in parity mode \cite{seedhouse2021pauli}, where we can distinguish if qubits involved in readout are projected to either the same or opposite state. Additional projection pulses can be used to recover full two-qubit correlations in this readout scheme \cite{philips2022universal}.

The included micromagnet produces a magnetic-field gradient along the qubit array, adding a contribution to the externally applied magnetic field and separating the resonant frequencies of the qubits. This allows for individual control of selected qubits using a shared gate while minimizing crosstalk. Experimental results of the frequency spacing are presented in Fig.\ref{fig:12Q}c. We find that at a field of 400 mT, the minimum frequency spacing between qubits is 14.8 MHz. We note a deviation of the expected frequency spacing from the experimental results. This deviation is a result of incomplete magnetization of the cobalt magnet which is strongly dependent on the microscopic details such as the size, position and orientation of each grain in vicinity of the qubits \cite{aldeghi2024simulation}. The simulations in Fig.\ref{fig:12Q}c assume a constant grain size of $\sim$14 nm in the micromagnet and is consistent with XRD measurements on blanket cobalt films, though we can expect a variability at the nanometer scale.

The coherent Rabi oscillations shown in Fig.\ref{fig:12Q}d can be performed individually on each of the 12 qubits by applying an oscillating signal at a given qubit's resonance frequency to the shared center screening gate CS, attributed to the magnetic-field gradient across the qubit channel to facilitate EDSR. With the same MW source output power we observe about +/-100 kHz deviation to the mean 600 kHz Rabi frequency (Fig.\ref{fig:12Q}e). The variation in Rabi frequencies is combination of (i) magnetic field gradient variations at the qubit regions (increasing by $\sim$10\% from Q1-Q8 and then decreasing from Q8-Q12 based on micromagnet simulations),  (ii) non ideal dot locations where few dots could be offset from the center of the qubit channel and the  (iii) MW transmission line response which can exhibit larger variation across a small frequency range resulting in various drive powers across the qubit channel.

The pure dephasing time T$_2^*$ time (with $\sim$5 min integration time) was 1.3 +/- 0.5 $\mu$s across the array (typ. 540 +/- 120 ns for 12-hour integration time on similar devices) with a corresponding Hahn echo time of T$_2^\text{H}$ = 112 +/- 13$\mu$s (Fig.\ref{fig:12Q}f).  In this sample the coherence times were limited by the nuclear-spin noise in the substrate due to the natural abundance of $^{29}$Si. Tunnel Falls devices with quantum wells of isotopically purified (800 ppm) $^{28}$Si have enhanced coherence times as expected \cite{neyens2024probing}.

Exchange coupling, which arises between spins in neighboring quantum dots, is used to implement two-qubit gates. While it is possible to operate with fixed exchange coupling \cite{huang2019fidelity}, higher control fidelities are reported in systems with this property being tunable \cite{mills2022two,xue2022quantum}. To probe the exchange coupling in the 12QD sample, we employ a spectroscopy measurement, where we observe splitting of the qubit EDSR resonance frequency as a function of the barrier voltage between itself and its neighbor (Fig.\ref{fig:12Q}g). We extract the tunability of the exchange coupling by calculating the amount of the voltage swing required on the barrier gate to achieve an order of magnitude increase in exchange coupling. We find that on average we need 35+/-7.2 mV/dec (Fig.\ref{fig:12Q}h), which is sufficient for more than three orders of magnitude exchange coupling tunability and should enable high-fidelity 2Q gates \cite{heinz2024analysis}. The variability of the exchange coupling tunability is strongly influenced by the device tuning, which can lead to quantum dots not being centered perfectly below their respective plunger gates. With an additional barrier balancing calibration, this variability can be reduced. Regardless, each exchange axis will be calibrated separately for high fidelity operations, so a small level of non-uniformity is acceptable, as long as each barrier allows for sufficient level of control.

In summary, Tunnel Falls spin-qubit devices are functional linear arrays of 12 QDs enabling up to 12 qubits which is the largest number of qubits reported in a single device in Si/SiGe \cite{philips2022universal}. Thousands of these devices are fabricated on each wafer using advanced 300 mm semiconductor manufacturing line and production-level process control. The work presented here validates the compatibility of high-volume manufacturing HVM fabrication to produce QD and spin-qubit devices using Si quantum wells in a SiGe heterostructure. Future work focuses on further material, fabrication process, and design optimization, as well as identifying process-qubit interactions that will help improve 1Q and 2Q fidelities. Results from this work provide the foundation for future spin-qubit devices and constitute an important step towards the realization of extensible 2-dimensional spin-qubit architectures. 

\textbf{REFERENCES: }
\bibliography{references}
\ \\
\noindent

\end{document}